\documentclass[10pt,journal]{IEEEtran}
\usepackage{amsmath,epsfig}
\usepackage{multirow}
\usepackage{float}
\usepackage{subfigure}
\usepackage{pgfplots}
\usepackage{pgfplotstable}
\usepackage{booktabs}
\usepackage{todonotes}
\usepackage{hyperref}
\usepackage{graphicx} 
\usepackage{breakurl}
\usepackage{caption}
\usepackage{tikz}
\usepackage{xcolor}
\usepackage{epstopdf}
\usepackage{amssymb}
\usepackage{soul,xcolor}

\usepackage{siunitx}

\usepackage{pifont}
\newcommand{\xmark}{\ding{55}}

\definecolor{compHR}{RGB}{133 ,   6,       10}
\definecolor{compMR}{RGB}{199,   14,        21}
\definecolor{compLR}{RGB}{243,   19,        29}

\definecolor{openHR}{RGB}{183 , 78 , 18}
\definecolor{openMR}{RGB}{244, 104 , 29}
\definecolor{openLR}{RGB}{ 247, 154 , 90}

\definecolor{light}{RGB}{255,   255,      0}
\definecolor{largeLow}{RGB}{192,   192,    192}
\definecolor{sparse}{RGB}{255,   204,    153}
\definecolor{industr}{RGB}{77,    77,     77}
\definecolor{denseTree}{RGB}{0,   102,      0}
\definecolor{scatTree}{RGB}{21,   255,     21}
\definecolor{bush}{RGB}{102,   153,      0}
\definecolor{lowPlant}{RGB}{204,   255,    102}
\definecolor{paved}{RGB}{0,     0,    102}
\definecolor{soil}{RGB}{   255,   255,    204}
\definecolor{water}{RGB}{ 110 ,107 , 254}

\makeatletter
\newenvironment{customlegend}[1][]{%
    \begingroup
    \let\addlegendimage=\pgfplots@addlegendimage
    \let\addlegendentry=\pgfplots@addlegendentry
    \pgfplots@init@cleared@structures
    \pgfplotsset{#1}%
}{%
    \pgfplots@createlegend
    \endgroup
}%
\makeatother

\usepackage[switch]{lineno} 
\usepackage{lipsum} 

\begin{document}
\title{Multi-level Feature Fusion-based CNN for Local Climate Zone Classification from Sentinel-2 Images: Benchmark Results on the So2Sat LCZ42 Dataset}

\author{Chunping Qiu, Xiaochong Tong, Michael Schmitt, \IEEEmembership{Senior Member, IEEE}, Benjamin Bechtel, Xiao Xiang Zhu \IEEEmembership{Senior Member, IEEE}

\IEEEcompsocitemizethanks{
C. Qiu and M. Schmitt are with Signal Processing in Earth Observation (SiPEO), Technical University of Munich (TUM), Germany. 
X. Tong is with Information Engineering University, Zhengzhou, China. 
Benjamin Bechtel is with the Institute of Geography, Ruhr-University Bochum, Germany. 
X. Zhu is with the Remote Sensing Technology Institute (IMF), German Aerospace Center (DLR), as well as TUM-SiPEO, Germany.
\emph{(Correspondence: Xiao Xiang Zhu; E-mail: xiaoxiang.zhu@dlr.de)}
}
}
 
\IEEEtitleabstractindextext{%
\begin{abstract}\textcolor{blue}{This article was accepted by IEEE Journal of Selected Topics in Applied Earth Observations and Remote Sensing.}
As a unique classification scheme for urban forms and functions, the local climate zone (LCZ) system provides essential general information for any studies related to urban \textcolor{black}{environments}, especially \textcolor{black}{on a large scale}. Remote sensing data-based classification approaches are the key to large-scale mapping and monitoring of LCZs. The potential of deep learning-based approaches is not yet fully explored, even though advanced convolutional neural networks (CNNs) continue to push the frontiers for various computer vision tasks. One reason is that published studies are based on different datasets, usually at \textcolor{black}{a} regional scale, which makes it impossible to fairly and consistently compare the potential of different CNNs for real-world scenarios. This study is based on the big So2Sat LCZ42 benchmark dataset dedicated \textcolor{black}{to} LCZ classification. Using this dataset, we studied a range of CNNs of varying sizes. In addition, we proposed a CNN to classify LCZs from Sentinel-2 images, Sen2LCZ-Net. Using this base network, we propose fusing multi-level features using the extended Sen2LCZ-Net-MF. With this proposed simple network architecture, and the highly competitive benchmark dataset, we obtain results that are better than those obtained by the state-of-the-art CNNs, \textcolor{black}{while} requiring less computation with fewer layers and parameters. Large-scale LCZ classification examples of completely unseen areas are presented, demonstrating the potential of our proposed Sen2LCZ-Net-MF as well as the So2Sat LCZ42 dataset. We also intensively investigated the influence of network depth and width, and the effectiveness of the design choices made for Sen2LCZ-Net-MF. Our work will provide important baselines for future CNN-based algorithm developments for \textcolor{black}{both} LCZ classification \textcolor{black}{and} other urban land cover land use classification. Code and pre-trained models are available at \url{https://github.com/ChunpingQiu/benchmark-on-So2SatLCZ42-dataset-a-simple-tour}.
\end{abstract}

}

\maketitle%
\IEEEdisplaynontitleabstractindextext
\IEEEpeerreviewmaketitle
\maketitle

\ifCLASSOPTIONcompsoc
    \IEEEraisesectionheading{\section{Introduction}\label{sec:introduction}}
\else
    \section{Introduction}
    \label{sec:intro}
\fi

\begin{figure*}[!tbh]
	\centering
    \includegraphics[width=0.99\textwidth]{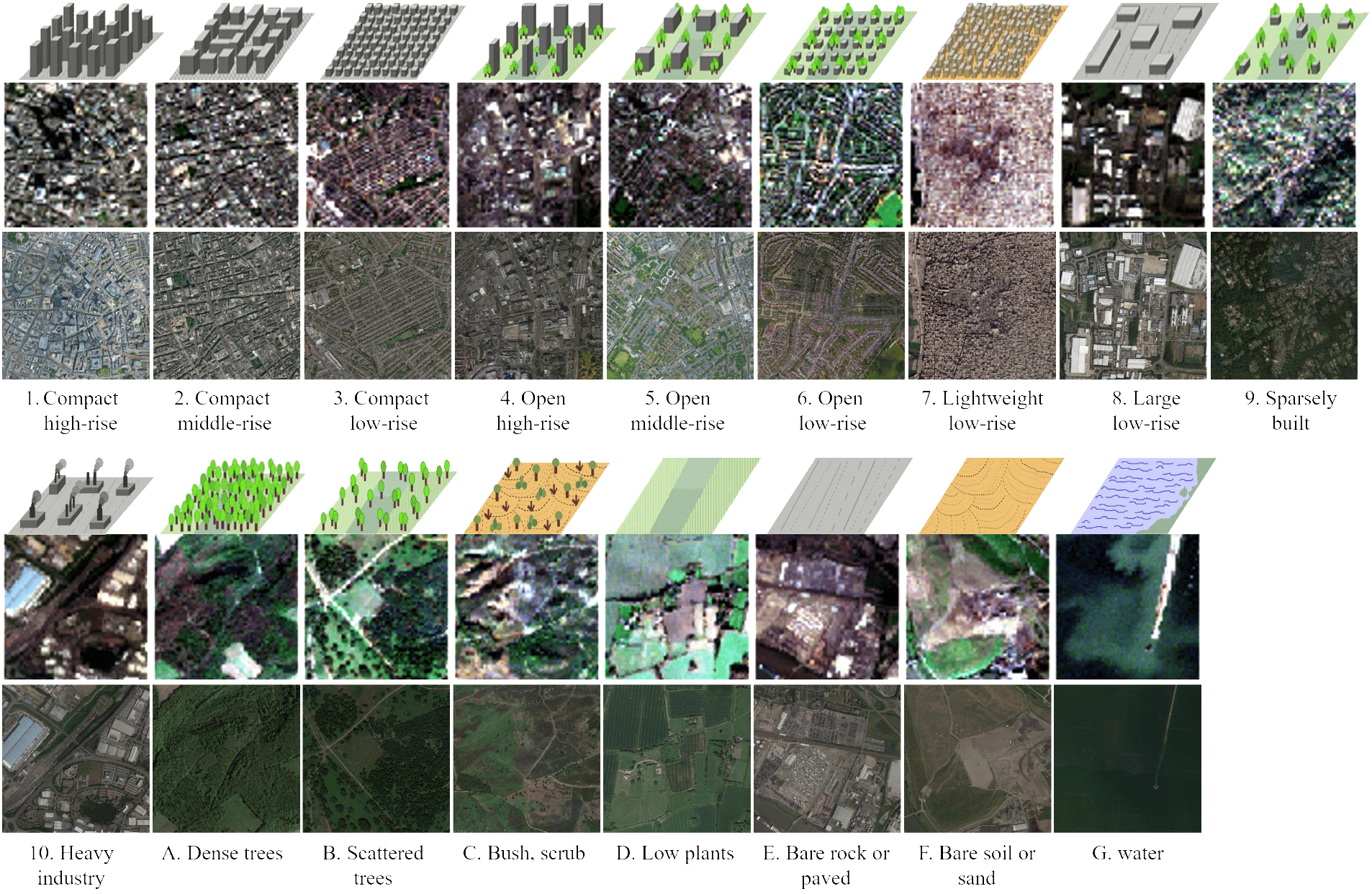}
	\caption{{Illustration of LCZs and corresponding Sentinel-2 and high resolution (HR) data patches. HR data source: Esri. Map image is the intellectual property of Esri and is used herein under license. Copyright \copyright 2019 Esri and its licensors. All rights reserved.} 
}
	\label{fig:lczS2HR}
\end{figure*}

The Local Climate Zone (LCZ) scheme is a classification system \textcolor{black}{that provides} a standardization framework for the characteristics of urban forms and functions. Illustrations of the LCZ classes and corresponding remote sensing image patches are shown in Fig.~\ref{fig:lczS2HR}. Originally proposed for urban heat island (UHI) research, this scheme has shown an increasing impact on various climatological studies, such as the cooling effect of green infrastructure and micro-climatic effects on town peripheries \cite{stewart2012local, stewart2014evaluation, quanz2018micro, kotharkar2018evaluating, geletivc2019inter, koc2018understanding, geletivc2019spatial}. Furthermore, the LCZ scheme can also be used to describe the internal structure of urban areas, providing significant information for various applications such as infrastructure planning and population assessment \cite{wicki2017attribution, ho2017spatial}.

An important part of the existing development of LCZ classification is community-based global LCZ mapping using openly available Landsat data and softwares \cite{mills2015introduction, bechtel2015mapping}. An example is the World Urban Database and Portal (WUDAPT) \cite{ching2018wudapt}, a community-driven initiative, which was organized by researchers to produce high-quality LCZ maps worldwide. Within WUDAPT, currently almost 100 cities located across the globe have been mapped \textcolor{black}{of} moderate quality, providing sufficient details for certain model applications \cite{bechtel2019generating}. LCZ maps of tens of cities, after quality assessment, are now openly available in the WUDAPT portal. More recently, an LCZ map of Europe was published \cite{demuzere2019mapping}.

The key to efficient large-scale LCZ classification is developing advanced machine learning models with high generalization ability \cite{demuzere2019mapping, MatthiasDemuzere.2019}. In this regard, tailoring deep learning-based approaches to the peculiarities of remote sensing data is one important strategy that \textcolor{black}{has} gained much attention recently \cite{xu2019urban, qiu2019local,qiu2019fusing,jing2019effective, yoo2019comparison,rosentreter2020towards,qiuFCN}. A review of these published studies tells us that deep learning, specifically in the form of convolutional neural networks (CNNs), is indeed able to enhance LCZ classification accuracy given a proper dataset due to its powerful feature representation \textcolor{black}{capacity}, when compared to random forest approaches \cite{zhu2017deep}. Specifically, some LCZs\textcolor{black}{,} such as open built-up areas and scattered trees, can benefit from the learned features that incorporate larger neighborhood information, as found in \cite{yoo2019comparison}.

However, while providing general meaningful insights into the methodology, most of the existing studies are carried out and assessed separately for individual case study scenes. This makes it difficult or impossible to compare different network architectures and \textcolor{black}{fairly} evaluate their \textcolor{black}{potential} for subsequent applications. More importantly, it hinders the development of more advanced approaches due to limited standard baselines. This dilemma is rooted in the fact that there exist only a few open datasets that are dedicated to large-scale LCZ classification. \textcolor{black}{Taking} the rapidly developing field of classification in computer vision as an example, based on the benchmark datasets such as ImageNet \cite{deng2009imagenet} and CIFAR \cite{krizhevsky2009learning}, research in CNNs has proliferated toward enhanced performance, simpler design, and higher efficiency\textcolor{black}{,} with new progress being achieved every year. Exemplary architectures include VGGNet \cite{simonyan2014very}, residual neural networks (ResNet) \cite{he2016deep}, densely connected convolutional networks (DenseNet) \cite{huang2017densely}, Inception \cite{szegedy2016rethinking}, and neural architecture search net (NASNet) \cite{zoph2016neural}. Aiming at improved performance, researchers have put significant effort into investigating the impact of pushing the depth \cite{he2016identity}, width \cite{zagoruyko2016wide}, and
cardinality \cite{xie2017aggregated} of networks. In addition to higher accuracy, strategies have also been developed in other dimensions, such as model simplicity and efficiency (e.g., efficient use of model parameters and the trade-off between accuracy and latency) \cite{howard2017mobilenets, chollet2017xception}, model scaling (with respect to depth/width/resolution) \cite{tan2019efficientnet}, and training strategy (e.g., hyper-parameters tuning and model parallelism) \cite{huang2019gpipe}. In particular, a great deal of recent effort has been \textcolor{black}{devoted to} designing efficiency, i.e., improving accuracy without hitting the hardware memory limit, e.g., by the increasingly popular neural architecture search (NAS) approach\cite{cai2018proxylessnas}. However, it is still an open research question whether those conclusions also hold for tasks in remote sensing, the confirmation of which requires a sequence of studies on well-designed benchmark datasets. Unfortunately, only \textcolor{black}{a} few benchmark datasets exist for satellite remote sensing, especially when it comes to medium-resolution images \cite{helber2019eurosat, schmitt2019sen12ms, sumbul2019bigearthnet}. All three existing datasets employ Sentinel-2 images and focus on applications such as land cover and land use classification. \textcolor{black}{One dataset}, \cite{ schmitt2019sen12ms} also includes Sentinel-1 images, providing the potential for multi-sensor fusion.

In the case of \textcolor{black}{the} LCZ classification scheme, the main reason \textcolor{black}{thre are so} few standard datasets is that it is costly, time-consuming, and challenging to collect ground truth for  classification schemes with so many difficult-to-distinguish classes. This problem is partly solved by the recently published, openly available So2Sat LCZ42 dataset, which contains both image patches and LCZ labels from 42 cities distributed across the world. Focusing on the specific task of LCZ classification under a specific setup with this dataset, a simple CNN architecture is proposed in this study. The influence of its depth, width, and pooling layers is extensively investigated. The best results are presented along with a wide range of baseline models of varying sizes, enabling a understanding of the correlation between model size and accuracy, and \textcolor{black}{supporting} further development \textcolor{black}{toward} higher classification accuracy.

The remainder of this paper proceeds as follows: Section~\ref{sec:method} elaborates \textcolor{black}{on} the proposed CNN architecture, Sen2LCZ-Net, and the strategy of multi-level feature fusion. Section~\ref{sec:exp} details descriptions \textcolor{black}{of} the So2Sat LCZ42 dataset, the baseline CNNs to be compared, and the experimental setup. Section~\ref{sec:res} evaluates the classification accuracy of LCZ and visualizes the classified LCZ results for several sample test scenes at both \textcolor{black}{a} city and province scale. The following Section~\ref{sec:dis} extensively investigates the influence and effect of the design choice of Sen2LCZ-Net-MF, and discusses multiple feasible approaches for further improvement\textcolor{black}{,} based on this study. Finally, Section~\ref{sec:con} summarizes and concludes the work.

\section{LCZ classification via CNNs}
\label{sec:method}

\subsection{An adapted CNN architecture and multi-level feature fusion}
\label{sec:methodProposed}

\textcolor{black}{Our priorities are} model simplicity and \textcolor{black}{the use of} fewer parameters\textcolor{black}{, which} benefit from the following two advantages. \textcolor{black}{First, a} simple, small model is more feasible for up-scaling LCZ classification---since generally, in big-data scenarios, a huge amount of data needs to be processed at a reasonable cost. \textcolor{black}{In addition, better and more discriminative features are encouraged to be learned by a small network with fewer trainable parameters \cite{hasanpour2016lets, hasanpour2018towards}, \textcolor{black}{thus} decreasing the chance of overfitting and enabling high generalization ability for LCZ classification \textcolor{black}{on a large scale}.}

\begin{figure*}[!tbh]
	\centering
    \includegraphics[width=0.99\textwidth]{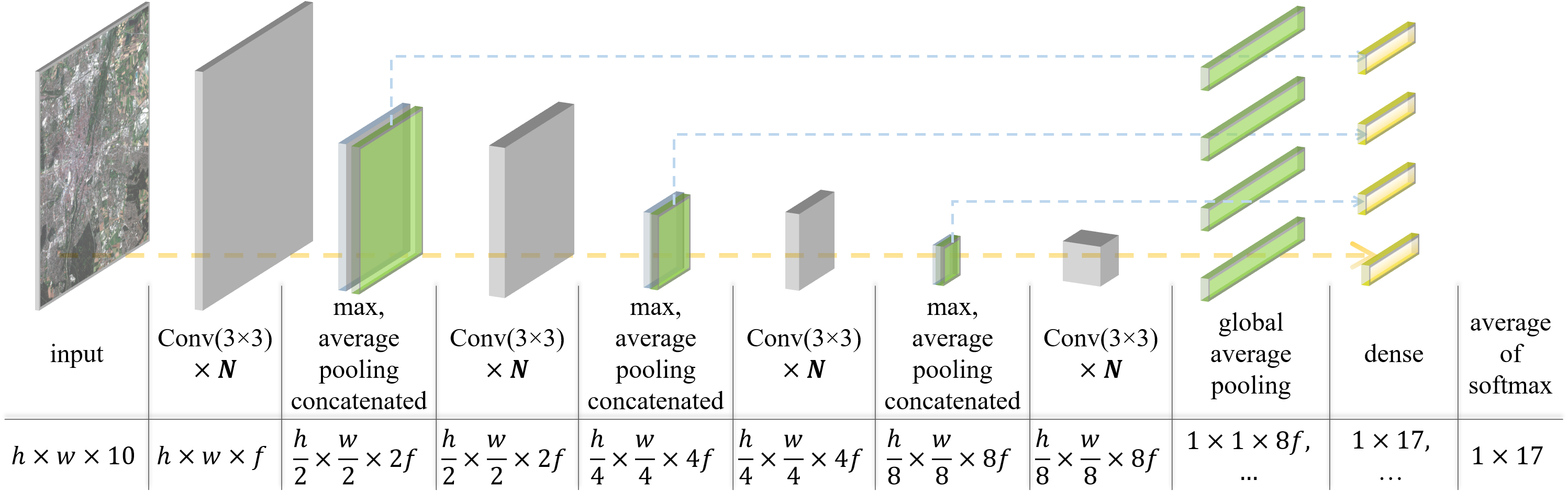}
	\caption{{Architecture of Sen2LCZ-MF. The three light blue lines correspond to the part of multi-level feature fusion. Note that all convolutional layers are followed by batch normalization. \textcolor{black}{The depth $D=4N+1$, and the width depends on the filter number of the first block\textcolor{black}{, which} is doubled for each subsequent block.}}
}
	\label{fig:neetwork}
\end{figure*}

Following the philosophy used by state-of-the-art models, \textcolor{black}{especially those aiming at simplicity}, our design {follows} the template \textcolor{black}{described in }\cite{springenberg2014striving, Goodfellow-et-al-2016}:
\begin{itemize}
\item Convolutional layers are with a fixed filter of a small size ($3\times 3$) for an efficient use of parameters;
\item Features maps are down-sampled to half the input resolution by using pooling layers\textcolor{black}{,} and the number of the computed feature maps is doubled to $2f$ after each down-sampling operation to enable hierarchical representation learning;
\item Homogeneous layers are grouped into blocks for network topology \textcolor{black}{to be easily managed}.
\end{itemize}
The proposed simple network architecture\textcolor{black}{,} that maps LCZs from Sentinel-2 by taking multi-level features into account, Sen2LCZ-Net-MF, is illustrated in Fig. \ref{fig:neetwork}, where detailed information (\textcolor{black}{layer names and sizes of feature maps}) about input, intermediate learned features, and output are also shown. Sen2LCZ-Net-MF consists of a simple end-to-end CNN, Sen2LCZ-Net, and connections to fuse multi-level features. In Sen2LCZ-Net, there are four sequential blocks that extract features via convolutional layers from the input patch or output of the previous block\textcolor{black}{; they then} abstract the learned features via average and maximum pooling layers, providing input for the subsequent block. The use of both average and maximum pooling layers within Sen2LCZ-Net, hereinafter referred to as the ``double-pooling layer,'' ensures that more learned features \textcolor{black}{or information within the input data} passes through \textcolor{black}{the network} for learning in a later stage. \textcolor{black}{This is especially important when the input patch size is small or has a coarse resolution, e.g., a $32\times 32$ Sentinel-2 patch.} At the end of the last block, a global average pooling is performed\textcolor{black}{,} followed by a softmax classifier for the final prediction. No additional fully connected layers are used\textcolor{black}{, in order to} maintain a small model size. 

The final prediction is then used for loss calculation and optimization\textcolor{black}{,} along with the reference label input as ground truth. When fusion of multi-level features is considered, four predictions are made from four outputs of the four blocks independently, the sum of which is used for loss calculation and optimization, as illustrated by the blue lines in Fig. \ref{fig:neetwork}. \textcolor{black}{Similar to the strategy using the double-pooling layer, this multi-level fusion design is intended to better exploit the information in input patches without introducing many more parameters. Low-level features from an early stage of the network can be valuable to distinguish LCZs such as sparsely built areas, while this information is not guaranteed to be available in the final learned, or high-level, features. It is worth mentioning that this fusion idea can be implemented together with any state-of-the-art CNNs, including VGG and ResNet. Furthermore, some state-of-the-art design ideas, such as attention mechanisms and skip connections, can be further integrated into Sen2LCZ-Net-MF; however, for simplicity, this is not recommended for consideration before its potential has been investigated.}

\textbf{Implementation Details.} 
Filter weights are initialized using the algorithm
proposed by \cite{he2015delving}. The kernel sizes of all convolutional layers are $3\times 3$ and during convolutions, each side of the inputs is zero-padded by one pixel to keep feature maps a fixed size. The number of output filters for the first convolutional layer, $f$, is set as 16, and the number of convolutional layers in each of the four blocks, $N$, is set as 4\textcolor{black}{;} experimented for investigations of the influence of network depth and width in Section \ref{sec:dis} experiment with the value of $N$. Changing the value of $f$ and $N$ results in different topologies of Sen2LCZ-Net. \textcolor{black}{The depth, $D$, and width of Sen2LCZ-MF can be adjusted with $N$ and $f$, respectively. Specifically, the depth $D=4N+1$, and the width $W$ depends on the filter number of the first block, $f$, which is doubled for each subsequent block.} The pooling layers use a kernel size of $2\times 2$ with a stride of 2, decreasing the size of feature maps by half. As a result, the sizes of the learned feature maps from the four blocks are $h\times w$, $\frac{h}{2}\times \frac{w}{2}$, $\frac{h}{4}\times \frac{w}{4}$, and $\frac{h}{8}\times \frac{w}{8}$. Specifically, on the So2Sat LCZ42 dataset, the sizes are $32\times 32$, $16\times 16$, $8\times 8$, and $4\times 4$. To avoid overfitting during training, we add a dropout layer \cite{srivastava2014dropout} \textcolor{black}{at} the end of the second and third block and set the dropout rate as 0.2.

\subsection{Baseline CNNs}
\label{sec:baselines}
To provide comparisons among baseline results, the following standard CNNs and modules were studied for LCZ classification. The baselines were selected to have a varying number of layers and parameters\textcolor{black}{,} to represent a wide range of cases.
\begin{itemize}
\item \textbf{VGG.} VGG is composed of convolutional layers with a fixed small size of $3\times 3$, max-pooling	layers with a size of $2\times 2$ and a stride of 2, and	three fully	connected layers	at	the	end  \cite{simonyan2014very}. When it was proposed, the authors showed that increasing the depth of the network is able to improve classification accuracy significantly (e.g., when pushing the depth from 16 to 19). Inspired by a earlier work \cite{zeiler2014visualizing}, it used only $3\times 3$ convolution filters to improve network performance and lower computational complexity. VGG became famous for its simplicity and homogeneous topology. However, one problem is the large \textcolor{black}{number} of parameters (mainly resulting from the three fully connected layers), making it computationally expensive. We used VGG16 implemented in Keras, with only one adaptation: adding batch normalization after each convolutional layer. The final size of the feature map for prediction is $1\times 4096$.
\item \textbf{ResNet.} 
ResNet improves training efficiency of very deep CNNs by introducing skip connections with identity functions. This design of short-cut connections helps to alleviate the vanishing gradients and diminishing feature reuse problems, enabling a very deep network\textcolor{black}{,} ranging from 152 to 1000 layers. In our comparative experiments, the {ResNet-11/20/54} used, as well as the attention-based ResNets are based on an improved version of the original proposed ResNet \cite{he2016identity, woo2018cbam}. There are three blocks in total, each of which outputs feature maps of size $32\times 32$, $16\times 16$, and $8\times 8$. 
The Keras implementation of ResNet50 used in this study is based on the first version \cite{he2016deep}, and the only adaptation \textcolor{black}{for} our experiments is the removal of the last block\textcolor{black}{,} due to the $32\times 32$ input patch size.
\item \textbf{ResNext.}  ResNext introduces cardinality as a new dimension of networks, in addition to depth and width \cite{xie2017aggregated}. It is similar to the family of Inception models in that they all follow the design strategy of split-transform-merge \cite{szegedy2015going}. That is, the input is first split into multiple embeddings (by $1\times1$ convolutions), then transformed by a set of specialized filters, and merged by concatenation or summation in the end. Improved classification accuracy has been obtained while the complexity is maintained, and no specialized design choices are needed due to the reduced number of hyper-parameters. The ResNext used in our experiments has a depth of 29 and a cardinality of 8.
\item \textbf{DenseNet.} Based on the success of ResNet, DenseNet further connects the layers within a network \cite{huang2017densely}. In DenseNet, each layer takes all preceding feature maps as input, and feature maps of each layer are used as inputs into all subsequent layers. In this way, the problem of vanishing-gradient is alleviated, the feature propagation is strengthened, and feature maps are better reused. 
The DenseNet used in our experiments has three dense blocks with an equal number of layers (7). The sizes of the feature maps from each block are $32\times 32$, $16\times 16$, and $8\times 8$. Following \cite{huang2017densely}, we also use $1\times 1$ convolution followed by $2\times 2$ average pooling as transition layers between two contiguous dense blocks. The hyper-parameter, growth rate, is set as 12.
\item \textbf{Xception.} The Xception network is an extreme version of Inception models that assume \textcolor{black}{that} cross-channel and spatial correlations can be mapped completely separately. It is built up by stacking depthwise separable filters (a replacement of the Inception modules in Inception models) with residual connections. This results in an efficient use of model parameters\textcolor{black}{,} as being shown in \cite{chollet2017xception}.  
We used Xception implemented in Keras, with no adaptation. The final size of the feature map for prediction is $1\times 2048$.
\item \textbf{Attention mechanism and Convolutional Block Attention Module (CBAM).} \textcolor{black}{Attention modules in the context of DL, originally popularized in the field of machine translation \cite{vaswani2017attention}, are capable of boosting the representation power of CNNs by integrating global contextual dependencies, in both the spatial and the spectral dimension. It works in a way similar to adaptative feature refinement and feature selection or recalibration in one \cite{hu2018squeeze} or more \cite{roy2018concurrent} dimensions of the feature maps.} The implementation of attention mechanism used in this study, CBAM, consists of a channel attention module and a spatial attention module in a sequential arrangement, with channel attention as the first \cite{woo2018cbam}. CBAM was used with ResNet-20 and ResNext in our experiments by appending one CBAM for each convoutional layer.
\end{itemize}

\textcolor{black}{In addition to the aforementioned baselines for comparison, we also investigated the use of CBAM and skip connections within the proposed Sen2LCZ-Net-MF. When using CBAM, it is appended to each convolutional layer without further adjustments. When skip connections are adopted, each block is treated independently with one shortcut connection. Full pre-activation is used for the skip connection in each block, i.e., output feature maps of the first convolutional layer are added to outputs of the last convolutional layer \cite{he2016identity}}

\subsection{LCZ classification Procedure}
\begin{figure*}[!tbh]
	\centering
    \includegraphics[width=0.8\textwidth]{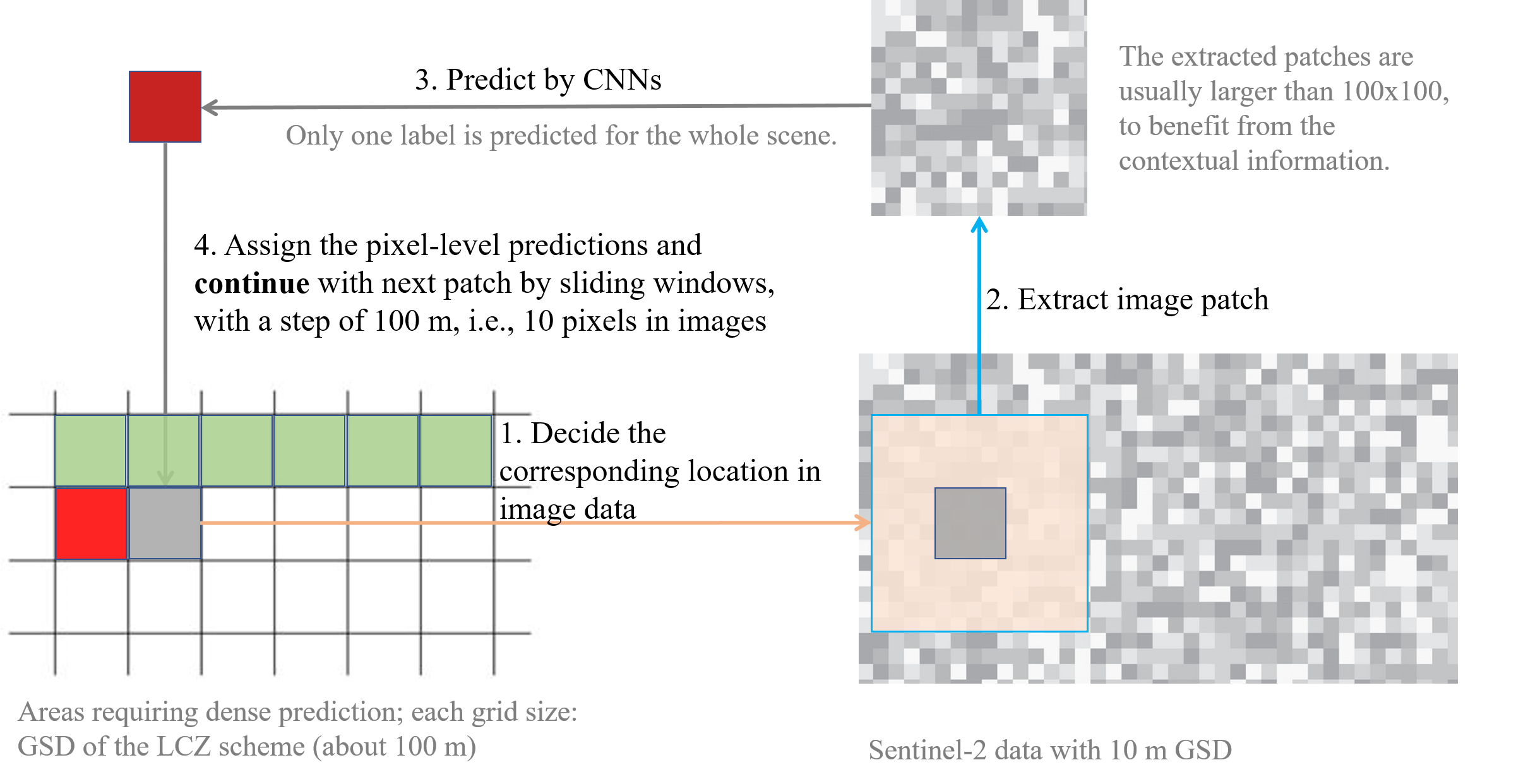}
	\caption{\textcolor{black}{Process of LCZ classification from image data by a sliding window approach with trained CNNs.}}
	\label{fig:mappingProces}
\end{figure*}

\textcolor{black}{After being trained, CNNs can be used for LCZ classification through a sliding window approach, as illustrated in Fig. \ref{fig:mappingProces}. For each pixel (location) to be predicted, one image patch is extracted from the whole image based on corresponding geo-locations, with a predefined size that is used during model training. Feeding the patch into trained CNNs will output a label corresponding to one of the 17 LCZs. The predicted label for this patch is assigned to this location before continuing to process the subsequent pixel (location). GSD of final predictions can be controlled via the step of the sliding window, which is often 100 m in LCZ-related studies.}

\section{Experimental Setup}
\label{sec:exp}

In this section, we describe the reference and image data and how they were processed, the baseline models, the experimental setup, and accuracy assessment.

\subsection{So2Sat LCZ42 dataset}

 The So2Sat LCZ42 dataset consists of LCZ labels of 400673 Sentinel-1 and Sentinel-2 image patches (\textcolor{black}{with a size of $32\times 32$}) in 42 urban agglomerations (plus 10 additional smaller areas) across the world. It was labeled by a group of domain experts in remote sensing, following a carefully designed labeling work flow similar to that in WUDAPT \cite{bechtel2019generating}. Afterwards, rigorous quality assessment was conducted with independent label voting by domain experts \textcolor{black}{who had} not labeled the areas in the labeling stage. The overall confidence of the labeling is 85\%. Further details about the So2Sat LCZ42 dataset can be found in \cite{zhu2019so2sat}. 
 
 In this study, only Sentinel-2 data was used. For the large-scale LCZ classification examples, the satellite imagery used was downloaded from Google Earth Engine after cloud removal processing, as described in \cite{Aggregating}. \textcolor{black}{Ten of the Sentinel-2 bands were used in this thesis: specifically, the channels with a GSD of 10 m and 20 m. In order to create composites with a consistent image size, the second group of bands was upsampled to a GSD of 10 m using cubic resampling. To summarize, the Sentinel-2 data used in this thesis contains 10 real-valued bands, as listed in Table \ref{tab:s2bands_40}.}

\begin{table}[!tbh]
  \centering
  \caption{Basic information of Sentinel-2 bands used in this study. VNIR: Visible and Near Infrared, SWIR: Short Wave Infrared}
    \begin{tabular}{llll}
    \toprule
    band & central wavelength (nm) & resolution (m) & description\\
    \midrule
    B2 & 490   & 10 & Blue\\
    B3 & 560   & 10 & Green\\
    B4 & 665   & 10 & Red\\
    B5 & 705   & 20, upsampled to 10 & VNIR\\
    B6 & 740   & 20, upsampled to 10 & VNIR \\
    B7 & 783   & 20, upsampled to 10 & VNIR\\
    B8 & 842   & 10                   & VNIR\\
    B8a & 865  & 20, upsampled to 10  & VNIR\\
    B11 &  1610& 20, upsampled to 10  & SWIR\\
    B12 &  2190& 20, upsampled to 10  & SWIR\\
    \bottomrule
    \end{tabular}
  \label{tab:s2bands_40}
\end{table}

 The whole So2Sat LCZ42 dataset is split into training, validation, and test sets, all of which are spatially separated. \textcolor{black}{Specifically, the training set consists of all the patches of 32 cities and the 10 add-on areas (see \cite{zhu2019so2sat} for the full list of cities). The remaining 10 cities, \textcolor{black}{distributed} across different continents over the world, are: Guangzhou, Jakarta, Moscow, Mumbai, Munich, Nairobi, San Francisco, Santiago de Chile, Sydney, and Tehran. For each of them, we split the labels of each LCZ class into the west and east halves of the city, comprising the validation and test sets. Therefore, all three sub-datasets are geographically separated from each other, even though the test and validation sets are drawn from the same list of cities.} The number and distribution of the individual LCZ classes in those three subsets is visualized in Fig. \ref{fig:disLCZ_dataset}. 
 
\pgfplotstableread{
label X1  X2 X3 X4  X5  X6 X7  X8 X9 X10  X11  X12 X13  X14 X15 X16  X17 num1  num2 num3 num4 num5 num6 num7  num8 num9 num10 num11 num12 num13  num14 num15 num16 num17
{Train}	1	2	3	4	5	6	7	8	9	10	11	12	13	14	15	16	17 5068	24431	31693	8651	16493	35290	3269	39326	13584	11954	42902	9514	9165	41377	2392	7898	49359
}\numTr

\pgfplotstableread{
label X1  X2 X3 X4  X5  X6 X7  X8 X9 X10  X11  X12 X13  X14 X15 X16  X17 num1  num2 num3 num4 num5 num6 num7  num8 num9 num10 num11 num12 num13  num14 num15 num16 num17
{va} 1	2	3	4	5	6	7	8	9	10	11	12	13	14	15	16	17	256	1254	2353 849	757	1906	474	3395	1914	860	2287	382	1202	2747 202	672	2609
}\numVa

\pgfplotstableread{
label X1  X2 X3 X4  X5  X6 X7  X8 X9 X10  X11  X12 X13  X14 X15 X16  X17 num1  num2 num3 num4 num5 num6 num7  num8 num9 num10 num11 num12 num13  num14 num15 num16 num17
{ts} 1	2	3	4	5	6	7	8	9	10	11	12	13	14	15	16	17	266	1262	2465	857	763	1936	503	3496	2013	908	2368	433	1166	2435	205	572	2540
}\numTs

\pgfplotsset{
    every tick/.style={very thin,gray},
    every tick label/.style={font={\scriptsize}},
    every axis label/.style={font={\small}},
    every axis/.append style={legend style={font=\tiny,line width=1pt,mark size=10pt}},
    ylabel near ticks, ylabel shift={-2pt}, xlabel shift={-1pt},
    }

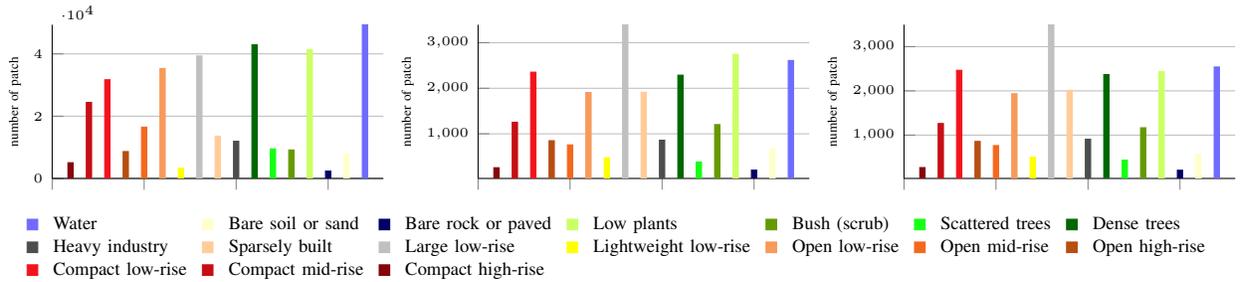
\begin{figure*}
\centering

\begin{tikzpicture}
\pgfplotsset{ybar,
    ymin=20,ymax=49359,xmin=0, xmax=18,
    height=0.2\linewidth,width=0.33\linewidth,
    every tick label/.style={font={\tiny}}}
\begin{axis}[
    name=plot1,
    bar width=2pt,
    xlabel={},
    xtick={},
    xticklabels from table={\numTr}{label},
    ylabel={\tiny number of patch},
    ymajorgrids,
    axis lines*=left]
  \addplot+ [color=compHR] table [y=num1, x=X1] \numTr;
\end{axis}
\begin{axis}[bar width=2pt,axis lines=none]
    \addplot+ [color=compMR] table [y=num2,x=X2] \numTr;
\end{axis}
\begin{axis}[bar width=2pt,axis lines=none]
    \addplot+ [color=compLR] table [y=num3,x=X3] \numTr;
\end{axis}
\begin{axis}[bar width=2pt,axis lines=none]
    \addplot+ [color=openHR] table [y=num4,x=X4] \numTr;
\end{axis}
\begin{axis}[bar width=2pt, axis lines=none]
    \addplot+ [color=openMR] table [y=num5,x=X5] \numTr;
\end{axis}
\begin{axis}[bar width=2pt,axis lines=none]
    \addplot+ [color=openLR] table [y=num6,x=X6] \numTr;
\end{axis}
\begin{axis}[bar width=2pt,axis lines=none]
    \addplot+ [color=light] table [y=num7,x=X7] \numTr;
\end{axis}
\begin{axis}[bar width=2pt,axis lines=none]
    \addplot+ [color=largeLow] table [y=num8,x=X8] \numTr;
\end{axis}
\begin{axis}[bar width=2pt,axis lines=none]
    \addplot+ [color=sparse] table [y=num9,x=X9] \numTr;
\end{axis}
\begin{axis}[bar width=2pt, axis lines=none]
    \addplot+ [color=industr] table [y=num10,x=X10] \numTr;
\end{axis}
\begin{axis}[bar width=2pt,axis lines=none]
    \addplot+ [color=denseTree] table [y=num11,x=X11] \numTr;
\end{axis}
\begin{axis}[bar width=2pt,axis lines=none]
    \addplot+ [color=scatTree] table [y=num12,x=X12] \numTr;
\end{axis}
\begin{axis}[bar width=2pt,axis lines=none]
    \addplot+ [color=bush] table [y=num13,x=X13] \numTr;
\end{axis}
\begin{axis}[bar width=2pt,axis lines=none]
    \addplot+ [color=lowPlant] table [y=num14,x=X14] \numTr;
\end{axis}
\begin{axis}[bar width=2pt,axis lines=none]
    \addplot+ [color=paved] table [y=num15,x=X15] \numTr;
\end{axis}
\begin{axis}[bar width=2pt, axis lines=none]
    \addplot+ [color=soil] table [y=num16,x=X16] \numTr;
\end{axis}
\begin{axis}[bar width=2pt,axis lines=none]
    \addplot+ [color=water] table [y=num17,x=X17] \numTr;
\end{axis}
\end{tikzpicture}
\begin{tikzpicture}
\pgfplotsset{ybar,
    ymin=20,ymax=3395,xmin=0, xmax=18,
    height=0.2\linewidth,width=0.33\linewidth,
    every tick label/.style={font={\tiny}}}
\begin{axis}[
    name=plot1,
    bar width=2pt,
    xlabel={},
    xtick={},
    xticklabels from table={\numVa}{label},
    ylabel={\tiny number of patch},
    ymajorgrids,
    axis lines*=left]
  \addplot+ [color=compHR] table [y=num1, x=X1] \numVa;
\end{axis}
\begin{axis}[bar width=2pt,axis lines=none]
    \addplot+ [color=compMR] table [y=num2,x=X2] \numVa;
\end{axis}
\begin{axis}[bar width=2pt,axis lines=none]
    \addplot+ [color=compLR] table [y=num3,x=X3] \numVa;
\end{axis}
\begin{axis}[bar width=2pt,axis lines=none]
    \addplot+ [color=openHR] table [y=num4,x=X4] \numVa;
\end{axis}
\begin{axis}[bar width=2pt, axis lines=none]
    \addplot+ [color=openMR] table [y=num5,x=X5] \numVa;
\end{axis}
\begin{axis}[bar width=2pt,axis lines=none]
    \addplot+ [color=openLR] table [y=num6,x=X6] \numVa;
\end{axis}
\begin{axis}[bar width=2pt,axis lines=none]
    \addplot+ [color=light] table [y=num7,x=X7] \numVa;
\end{axis}
\begin{axis}[bar width=2pt,axis lines=none]
    \addplot+ [color=largeLow] table [y=num8,x=X8] \numVa;
\end{axis}
\begin{axis}[bar width=2pt,axis lines=none]
    \addplot+ [color=sparse] table [y=num9,x=X9] \numVa;
\end{axis}
\begin{axis}[bar width=2pt, axis lines=none]
    \addplot+ [color=industr] table [y=num10,x=X10] \numVa;
\end{axis}
\begin{axis}[bar width=2pt,axis lines=none]
    \addplot+ [color=denseTree] table [y=num11,x=X11] \numVa;
\end{axis}
\begin{axis}[bar width=2pt,axis lines=none]
    \addplot+ [color=scatTree] table [y=num12,x=X12] \numVa;
\end{axis}
\begin{axis}[bar width=2pt,axis lines=none]
    \addplot+ [color=bush] table [y=num13,x=X13] \numVa;
\end{axis}
\begin{axis}[bar width=2pt,axis lines=none]
    \addplot+ [color=lowPlant] table [y=num14,x=X14] \numVa;
\end{axis}
\begin{axis}[bar width=2pt,axis lines=none]
    \addplot+ [color=paved] table [y=num15,x=X15] \numVa;
\end{axis}
\begin{axis}[bar width=2pt, axis lines=none]
    \addplot+ [color=soil] table [y=num16,x=X16] \numVa;
\end{axis}
\begin{axis}[bar width=2pt,axis lines=none]
    \addplot+ [color=water] table [y=num17,x=X17] \numVa;
\end{axis}
\end{tikzpicture}
\begin{tikzpicture}
\pgfplotsset{ybar,
    ymin=20,ymax=3496,xmin=0, xmax=18,
    height=0.2\linewidth,width=0.33\linewidth,
    every tick label/.style={font={\tiny}}}
\begin{axis}[
    name=plot1,
    bar width=2pt,
    xlabel={},
    xtick={},
    xticklabels from table={\numTs}{label},
    ylabel={\tiny number of patch},
    ymajorgrids,
    axis lines*=left]
  \addplot+ [color=compHR] table [y=num1, x=X1] \numTs;
\end{axis}
\begin{axis}[bar width=2pt,axis lines=none]
    \addplot+ [color=compMR] table [y=num2,x=X2] \numTs;
\end{axis}
\begin{axis}[bar width=2pt,axis lines=none]
    \addplot+ [color=compLR] table [y=num3,x=X3] \numTs;
\end{axis}
\begin{axis}[bar width=2pt,axis lines=none]
    \addplot+ [color=openHR] table [y=num4,x=X4] \numTs;
\end{axis}
\begin{axis}[bar width=2pt, axis lines=none]
    \addplot+ [color=openMR] table [y=num5,x=X5] \numTs;
\end{axis}
\begin{axis}[bar width=2pt,axis lines=none]
    \addplot+ [color=openLR] table [y=num6,x=X6] \numTs;
\end{axis}
\begin{axis}[bar width=2pt,axis lines=none]
    \addplot+ [color=light] table [y=num7,x=X7] \numTs;
\end{axis}
\begin{axis}[bar width=2pt,axis lines=none]
    \addplot+ [color=largeLow] table [y=num8,x=X8] \numTs;
\end{axis}
\begin{axis}[bar width=2pt,axis lines=none]
    \addplot+ [color=sparse] table [y=num9,x=X9] \numTs;
\end{axis}
\begin{axis}[bar width=2pt, axis lines=none]
    \addplot+ [color=industr] table [y=num10,x=X10] \numTs;
\end{axis}
\begin{axis}[bar width=2pt,axis lines=none]
    \addplot+ [color=denseTree] table [y=num11,x=X11] \numTs;
\end{axis}
\begin{axis}[bar width=2pt,axis lines=none]
    \addplot+ [color=scatTree] table [y=num12,x=X12] \numTs;
\end{axis}
\begin{axis}[bar width=2pt,axis lines=none]
    \addplot+ [color=bush] table [y=num13,x=X13] \numTs;
\end{axis}
\begin{axis}[bar width=2pt,axis lines=none]
    \addplot+ [color=lowPlant] table [y=num14,x=X14] \numTs;
\end{axis}
\begin{axis}[bar width=2pt,axis lines=none]
    \addplot+ [color=paved] table [y=num15,x=X15] \numTs;
\end{axis}
\begin{axis}[bar width=2pt, axis lines=none]
    \addplot+ [color=soil] table [y=num16,x=X16] \numTs;
\end{axis}
\begin{axis}[bar width=2pt,axis lines=none]
    \addplot+ [color=water] table [y=num17,x=X17] \numTs;
\end{axis}
\end{tikzpicture}
    
    \begin{tikzpicture}
    \pgfplotsset{
    legend style={cells={anchor=west}, draw=none,column sep=1ex,     nodes={scale=0.7, transform shape}}
    }
        \begin{customlegend}[legend columns=7]
        \addlegendimage{water, only marks, mark=square*}
        \addlegendentry{Water}
        \addlegendimage{soil, only marks, mark=square*}
        \addlegendentry{Bare soil or sand}
        \addlegendimage{paved, only marks, mark=square*}
        \addlegendentry{Bare rock or paved}
        \addlegendimage{lowPlant, only marks, mark=square*}
        \addlegendentry{Low plants}
        \addlegendimage{bush, only marks, mark=square*}
        \addlegendentry{Bush (scrub)}
        \addlegendimage{scatTree, only marks, mark=square*}
        \addlegendentry{Scattered trees}
        \addlegendimage{denseTree, only marks, mark=square*}
        \addlegendentry{Dense trees}
        \addlegendimage{industr, only marks, mark=square*}
        \addlegendentry{Heavy industry}
        \addlegendimage{sparse, only marks, mark=square*}
        \addlegendentry{Sparsely built}

        \addlegendimage{largeLow, only marks, mark=square*}
        \addlegendentry{Large low-rise}
        \addlegendimage{light, only marks, mark=square*}
        \addlegendentry{Lightweight low-rise}        
        \addlegendimage{openLR, only marks, mark=square*}
        \addlegendentry{Open low-rise}
        \addlegendimage{openMR, only marks, mark=square*}
        \addlegendentry{Open mid-rise}
        \addlegendimage{openHR, only marks, mark=square*}
        \addlegendentry{Open high-rise}
        \addlegendimage{compLR, only marks, mark=square*}
        \addlegendentry{Compact low-rise}
        \addlegendimage{compMR, only marks, mark=square*}
        \addlegendentry{Compact mid-rise}
        \addlegendimage{compHR, only marks, mark=square*}
\addlegendentry{Compact high-rise}
\end{customlegend}
\end{tikzpicture}
   \caption{Sample number of LCZ labels in training (left), validation (middle), and test datasets (right).}
\label{fig:disLCZ_dataset}
\end{figure*}

\subsection{Experimental settings and metrics for accuracy assessment}

\textbf{Training.} For all the CNNs studied in our experiments, the input images and their corresponding reference labels are used to train the network with the Nesterov Adam optimizer implementation of Keras \cite{chollet2015keras}. \textcolor{black}{All CNNs were trained from scratch following the same experimental settings in order to make meaningful comparisons.} We used a minibatch size of 32 patches. The initial learning rate is $2\times10^{-2}$ and is decreased by half after every fifth epoch. To control the training time and avoid overfitting, early stopping was used, and the monitored metric is validation loss with patience of 40 epochs, which means that the training stops if the validation loss does not decrease for 40 epochs. After the training, we report the test accuracy from the saved weights with the highest validation accuracy.

\textbf{Metrics.} Metrics used for performance assessment include overall accuracy (OA), Kappa, and average accuracy (AA)\textcolor{black}{,} which is chosen considering the unbalanced number of samples of different LCZs \cite{benediktsson2015spectral}. Additionally, we used weighted accuracy (WA), in which different weights are given to different types of mistakes on the basis of a \textcolor{black}{systematic} analysis of the consequent climate impact of those misclassfications, considering \textcolor{black}{such} properties as openness, height, cover, and thermal inertia \cite{bechtel2017quality}. As in \cite{yoo2019comparison}, overall accuracies for LCZ types in built-up areas (i.e., LCZ1-10; OA\_b) and LCZ types in non-built-up areas (i.e., LCZA-G; OA\_nb) are also used as auxiliary metrics.


\section{Experimental Results}
\label{sec:res}

Results of the experimental assessment of the proposed Sen2LCZ-Net-MF are given in this section. First, sensitivity analyses are carried out by comparing the performance of Sen2LCZ-Net-MF with different configurations, in order to back up the design choice and also to search for the best configuration. Using the chosen configuration, LCZ classification results are then compared to those from the state-of-the-art CNNs. 
Additionally, we show the detailed class-wise classification accuracy and remaining confusions among LCZs by a confusion matrix. In the end, we show the generally satisfying LCZ classification results \textcolor{black}{on a large scale}, with examples from two cities and a province under different environmental conditions.

\subsection{Influence of network depth and width}

It is well known that the depth and width of a CNN affect its performance. Specifically, bigger models tend to achieve higher accuracy and the accuracy gain quickly saturates \cite{tan2019efficientnet}. There are also contradictory observations where a deeper CNN does not achieve as
good accuracy as a shallower counterpart with the same number of parameters \cite{hasanpour2018towards}. 
To study the influence of depth and width of the Sen2LCZ-Net on the So2Sat LCZ42 dataset, we carried out a series of experiments following the same setup. The results are presented in Table \ref{tab:networkShapeEffect}, where we also list the number of trainable parameters for each Sen2LCZ-Net.

\textcolor{black}{From Table \ref{tab:networkShapeEffect}, we can observe the following phenomena:}
\begin{itemize}
\item When the number of feature maps in the first block, $f$, is set to 16, better classification results can be achieved as the network depth increases from 5 to 9, from 9 to 13, \textcolor{black}{and} from 13 to 17. The improvement from 13 to 17 is smaller than that from 5 to 9 and from 9 to 13. No further improvement is observed when the depth continues to increase from 17 to 21.
\item When $f$ is 32, unexpectedly, for the same depth (5, 9, and 17), a wider Sen2LCZ-Net does not provide obvious benefit, even though many more parameters are used. One explanation is that the bigger CNNs with more parameters tend to overfit on the training data, resulting in low test accuracy.
\item \textcolor{black}{A} correlation between model performance and size on the So2Sat LCZ42 dataset, i.e., larger models tend to demonstrate better performance until a certain threshold, is consistent with the literature \cite{tan2019efficientnet}.
\item With similar amount of parameters, e.g., f16D17 and f32D5, a deeper network provides better performance, which is probably due to the over saturation of the parameters in the shallow network (a phenomena called processing level saturation) \cite{hasanpour2018towards}.
\item A deeper network with fewer parameters (e.g., {f16D13} and {f16D9}) can perform better than its shallower counterparts (e.g., {f32D5}), probably due to its developed composition of more general simple functions.
\end{itemize}
\begin{table*}
  \centering
  \caption{Test results from Sen2LCZ-Net of different depth (number of layer, $D=4N+1$) and width\textcolor{black}{, which} is related to the filter number of the first block, $f$.}
    \begin{tabular}{ccccccccccc}
    \toprule
    \multicolumn{4}{c}{networks}  &       & \multicolumn{6}{c}{metrics} \\
\cmidrule{1-4}\cmidrule{6-11}    f     & D & \# Para. & ID    &       & Kappa & AA    & WA    & OA    & OA\_b & OA\_nb \\
    \midrule
    \multirow{4}[1]{*}{16} & 5     & 197,889 & f16D5 &       & 0.611 & 0.526 & 0.911 & 0.645 & 0.910 & 0.971 \\
          & 9     & 394,449 & f16D9 &       & 0.627 & 0.535 & 0.920 & 0.660 & 0.958 & 0.953 \\
          & 13    & 591,009 & F16-L13 &       & 0.636 & \textbf{0.557} & 0.924 & 0.668 & 0.947 & \textbf{0.975} \\
          & 17    & 787,569 & f16D17 &       & \textbf{0.646} & {0.554} & \textbf{0.932} & \textbf{0.677} & \textbf{0.983} & 0.944 \\
          & 21    & 984,129 & F16-L21 &       & 0.632 & 0.549 & 0.925 & 0.664 & 0.963 & 0.954 \\
\cmidrule{2-11}    \multirow{3}[2]{*}{32} & 5     & 782,833 & f32D5 &       & 0.594 & 0.533 & 0.915 & 0.628 & 0.942 & 0.960 \\
          & 9     & 1,567,633 & f32D9 &       & 0.608 & 0.523 & 0.920 & 0.642 & 0.959 & 0.973 \\
          & 17    & 3,137,233 & F32-L17 &       & 0.644 & 0.552 & 0.927 & 0.676 & 0.955 & 0.969 \\
    \bottomrule
    \end{tabular}%
  \label{tab:networkShapeEffect}%
\end{table*}%

\subsection{Effectiveness of multi-level feature fusion}

To demonstrate the effectiveness of multi-level feature fusion for LCZ classification, we carried out 12 experiments with six CNNs with and without multi-level feature fusion. The resulting model performance is shown in Table \ref{tab:mfEffect}. Improvement from multi-level fusion can be consistently observed for all six CNNs of varying size. This improvement is more likely resulting from the representation ability of the additional employed features, since there are only a few additionally introduced parameters, i.e., those from the additional three dense and softmax layers,\textcolor{black}{,} as shown in Fig. \ref{fig:neetwork}. Another related explanation is that the additionally utilized feature maps from the early layers are larger and provide valuable information, enabling an efficient harness of the training samples, as analyzed in \cite{hasanpour2016lets}.

\begin{table}
  \centering
  \caption{Testing performance of six CNNs with (\checkmark) and without multi-level fusion (MF).}
    \begin{tabular}{cccccccc}
    \toprule
    \multirow{2}[4]{*}{network} & \multirow{2}[4]{*}{MF} & \multicolumn{6}{c}{metrics} \\
\cmidrule{3-8}          &       & Kappa & AA   & WA    & OA    & OA\_b & OA\_nb \\
    \midrule
    \multirow{2}[2]{*}{f16D5} &       & 0.611 & 0.526 & 0.911 & 0.645 & 0.910 & \textbf{0.971} \\
          & \checkmark & \textbf{0.628} & \textbf{0.528} & \textbf{0.920} & \textbf{0.661} & \textbf{0.949} & 0.959 \\
    \midrule
    \multirow{2}[2]{*}{f16D9} &       & 0.627 & 0.535 & 0.920 & 0.660 & \textbf{0.958} & 0.953 \\
          & \checkmark & \textbf{0.644} & \textbf{0.562} & \textbf{0.928} & \textbf{0.674} & 0.957 & \textbf{0.967} \\
    \midrule
    \multirow{2}[2]{*}{f16D17} &       & 0.646 & 0.554 & 0.932 & 0.677 & \textbf{0.983} & 0.944 \\
          & \checkmark & \textbf{0.664} & \textbf{0.587} & \textbf{0.933} & \textbf{0.694} & 0.973 & \textbf{0.959} \\
    \midrule
    \multirow{2}[2]{*}{f32D5} &       & 0.594 & 0.533 & 0.915 & 0.628 & 0.942 & \textbf{0.960} \\
          & \checkmark & \textbf{0.633} & \textbf{0.546} & \textbf{0.923} & \textbf{0.666} & \textbf{0.957} & 0.957 \\
    \midrule
    \multirow{2}[2]{*}{f32D9} &       & 0.608 & 0.523 & \textbf{0.920} & 0.642 & \textbf{0.959} & \textbf{0.973} \\
          & \checkmark & \textbf{0.635} & \textbf{0.554} & \textbf{0.920} & \textbf{0.668} & 0.954 & \textbf{0.966} \\
    \midrule
    \multirow{2}[2]{*}{ResNet-54} &       & 0.559 & 0.481 & 0.894 & 0.597 & 0.876 & \textbf{0.952} \\
          & \checkmark & \textbf{0.581} & \textbf{0.483} & \textbf{0.907} & \textbf{0.619} & \textbf{0.943} & 0.944 \\
    \bottomrule
    \end{tabular}%
  \label{tab:mfEffect}%
\end{table}%

\subsection{Effectiveness of the double-pooling layer}

It is known that (max-)pooling plays an important role in reducing the sensitivity of learned features to shift and distortions and that it also enables translation invariance \cite{scherer2010evaluation}. Also, it makes possible a pyramid-shaped form model and larger receptive field in a later stage of the network. However, it hinders the maximum utilization of information, which is crucial, especially for a dataset \textcolor{black}{of insufficient size}. Therefore, both average pooling and maximum pooling layers are used for downsampling within Sen2LCZ-Net. To show the effectiveness of this choice, we compared the results to those from models only using maximum pooling layers for six configurations. The comparisons are presented in Table \ref{tab:amEffect}, where it can be seen that it is beneficial to use average pooling in addition to maximum pooling layers. This is probably because more features and information can be exploited for a later stage of the network by simply adding average pooling. This is important for the used medium resolution Sentinel-2 images\textcolor{black}{,} because \textcolor{black}{each} pixel value represents a rather large ground area and might be crucial \textcolor{black}{in} distinguish certain LCZs \textcolor{black}{that often depend on neighborhood morphologies}. When only maximum pooling layers are used, certain information can be lost during the feature extraction and abstraction process, and \textcolor{black}{cannot} be recovered.

\begin{table*}[htbp]
  \centering
  \caption{Testing performance of six CNNs with (\checkmark) and without double-pooling layer. AM indicates the configuration where both pooling layers, average and maximum pooling, are used. Without double-pooling layer means only maximum pooling is used.}
    \begin{tabular}{cccccccc}
    \toprule
    \multirow{2}[4]{*}{network} & \multirow{2}[4]{*}{AM} & \multicolumn{6}{c}{metrics} \\
\cmidrule{3-8}          &       & Kappa & AA    & WA    & OA    & OA\_b & OA\_nb \\
    \midrule
    \multirow{2}[2]{*}{f16D5} &       & 0.577 & 0.514 & 0.909 & 0.613 &\textbf{ 0.964} & 0.924 \\
          & \checkmark & \textbf{0.611} & \textbf{0.526} & \textbf{0.911} & \textbf{0.645} & 0.910 & \textbf{0.971} \\
    \midrule
    \multirow{2}[2]{*}{f16D9} &       & 0.611 & 0.534 & \textbf{0.920} & 0.644 & 0.944 & \textbf{0.972} \\
          & \checkmark & \textbf{0.627} & \textbf{0.535} & \textbf{0.920} & \textbf{0.660} & \textbf{0.958} & 0.953 \\
    \midrule
    \multirow{2}[2]{*}{f16D17} &       & \textbf{0.651} & \textbf{0.565} & \textbf{0.932} & \textbf{0.682} & 0.976 & \textbf{0.949} \\
          & \checkmark & 0.646 & 0.554 & \textbf{0.932} & 0.677 & \textbf{0.983} & 0.944 \\
    \midrule
    \multirow{2}[2]{*}{f16D5-MF} &       & 0.622 & 0.524 & 0.919 & 0.656 & \textbf{0.949} & \textbf{0.960} \\
          & \checkmark & \textbf{0.628} & \textbf{0.528} & \textbf{0.920} & \textbf{0.661} & \textbf{0.949} & 0.959 \\
    \midrule
    \multirow{2}[2]{*}{f16D9-MF} &       & 0.638 & 0.557 & 0.925 & 0.670 & 0.951 & 0.960 \\
          & \checkmark & \textbf{0.644} & \textbf{0.562} & \textbf{0.928} & \textbf{0.674} & \textbf{0.957} & \textbf{0.967} \\
    \midrule
    \multirow{2}[2]{*}{f16D17-MF} &       & 0.648 & 0.561 & 0.929 & 0.679 & 0.964 & \textbf{0.976} \\
          & \checkmark & \textbf{0.664} & \textbf{0.587} & \textbf{0.933} & \textbf{0.694} & \textbf{0.973} & 0.959 \\
    \bottomrule
    \end{tabular}%
  \label{tab:amEffect}%
\end{table*}%

\subsection{Impact of LCZ imbalance in the training dataset}

Samples in the training dataset are unbalanced with respect to each LCZ, as can be seen in Fig. \ref{fig:disLCZ_dataset}. To understand its effect on classification performance, we carried out eight experiments with four CNNs and compared the results considering and not considering the class imbalance problem. 
Class weights were used when considering the class imbalance problem, and weights were calculated based on the sample frequency of each LCZ. \textcolor{black}{Specifically, the weight of each LCZ was calculated by the inverse of its sample fraction in the whole training set.} The resulting differences in classification accuracy can be seen in Table \ref{tab:weightTrain}, Counterintuitively, we do not observe obvious benefits by using class weights. One reason might be that the imbalance problem in the So2Sat LCZ42 dataset is not serious\textcolor{black}{,} so that no weighting strategy is needed, as the imbalance problem was addressed to some extent during the data preparation process \cite{zhu2019so2sat}. Another reason is that the introduced class weight during training makes it difficult to learn generalized features for the major class, leading to a overall worse results.

\begin{table}[htbp]
  \centering
  \caption{Testing performance of four CNNs trained with and without (\xmark) class weights.}
    \begin{tabular}{cccccccc}
    \toprule
    \multirow{2}[4]{*}{network} & \multirow{2}[4]{*}{W} & \multicolumn{6}{c}{metrics} \\
\cmidrule{3-8}          &       & Kappa & AA   & WA    & OA    & OA\_b & OA\_nb \\
    \midrule
    \multirow{2}[2]{*}{f16D9-MF} &       & 0.591 & 0.555 & 0.908 & 0.623 & 0.944 & 0.958 \\
          &    \xmark   & \textbf{0.644} & \textbf{0.562} & \textbf{0.928} & \textbf{0.674} & \textbf{0.957} & \textbf{0.967} \\
    \midrule
    \multirow{2}[2]{*}{f16D17-MF} &       & 0.619 & 0.583 & 0.921 & 0.650 & 0.953 & \textbf{0.959} \\
          &   \xmark    & \textbf{0.664} & \textbf{0.587} & \textbf{0.933} & \textbf{0.694} & \textbf{0.973} & \textbf{0.959} \\
    \midrule
    \multirow{2}[2]{*}{ResNet-11} &       & 0.584 & 0.548 & 0.904 & 0.619 & 0.925 & 0.952 \\
          &    \xmark   & \textbf{0.620} & \textbf{0.551} & \textbf{0.918} & \textbf{0.654} & \textbf{0.943} & \textbf{0.963} \\
    \midrule
    \multirow{2}[2]{*}{ResNet-20} &       & 0.608 & \textbf{0.564} & \textbf{0.919} & 0.641 & 0.932 & \textbf{0.979} \\
          &    \xmark   & \textbf{0.609} & 0.527 & 0.918 & \textbf{0.642} & \textbf{0.960} & 0.957 \\
    \bottomrule
    \end{tabular}%
  \label{tab:weightTrain}%
\end{table}%

\subsection{\textcolor{black}{Comparison among state-of-the-art CNNs} for LCZ classification}
Table \ref{tab:acc} presents classification results from Sen2LCZ-Net-MF with selected configurations based on the above analyses, as well as several baseline CNNs, as described in Section \ref{sec:baselines}. The proposed Sen2LCZ-Net-MF(f16D17), corresponding to a configuration of $f=16$ and $D=4N+1, N=4$ \textcolor{black}{as described in Section \ref{sec:methodProposed} and Fig. \ref{fig:neetwork}},
provides the best results for all metrics except OA\_nb, with Kappa and AA being 0.664 and 0.587, respectively. A smaller version of Sen2LCZ-Net-MF(f9D17) also provides top results for all metrics, with Kappa and AA being 0.644 and 0.562, respectively. CNNs providing comparative results include ResNet-11, DensNet, and CBAM-based ResNet-20. Considering that the top results of OA\_b and OA\_nb are all close and high, it can be concluded that the best results are from the proposed Sen2LCZ-Net-MF. The bigger models, such as ResNet-50 and Xception, provides worse results. The first reason is that they are probably not adapted well for this task and dataset. Specifically, the information loss in the first layers of these CNNs, which is due to maximum pooling layers and the larger-than-1 stride of convolutional layers, might be harmful for feature representations.

\textcolor{black}{Further exploiting CBAM or skip connections provides little improvement, as presented in the last two rows in Table \ref{tab:acc}. This is possibly due to the sufficient exploitation of the input data by Sen2LCZ-Net-MF and the challenges of distinguishing different LCZs.}

It should be mentioned that the comparison is mainly to provide a reference for a preliminary interpretation of the relation between the models and their performance for the specific task of LCZ classification. It becomes clear that optimal performance for LCZ classification is not guaranteed when relying on the models proposed for datasets in computer vision\textcolor{black}{,} such as ImageNet.

\begin{table*}
  \centering
  \caption{Performance comparison among various CNNs on So2Sat LCZ42 dataset, following the same experimental setup. The top five results are indicated in bold.}
    \begin{tabular}{llccccccc}
    \toprule
    \multicolumn{2}{c}{CNNs} &       & \multicolumn{6}{c}{metrics} \\
\cmidrule{1-2}\cmidrule{4-9}    name  & \# Para. &       & Kappa & AA    & WA    & OA    & OA\_b & OA\_nb \\
    \midrule
    Sen2LCZ-Net-MF(F16-L17) & 791,428 &       & \textbf{0.664} & \textbf{0.587} & \textbf{0.933} & \textbf{0.694} & \textbf{0.973} & 0.959 \\
    Sen2LCZ-Net-MF(F16-L9) & 398,308 &       & \textbf{0.644} & \textbf{0.562} & \textbf{0.928} & \textbf{0.674} & 0.957 & \textbf{0.967} \\
    \midrule
    ResNet-11 & 300,225 &       & 0.620 & \textbf{0.551} & 0.918 & 0.654 & 0.943 & \textbf{0.963} \\
    DensNet & 389,189 &       & \textbf{0.631} & 0.550 & \textbf{0.924} & \textbf{0.663} & \textbf{0.964} & \textbf{0.962} \\
    ResNet-20 & 573,409 &       & 0.599 & 0.511 & 0.917 & 0.635 & 0.955 & \textbf{0.968} \\
    ResNet-20+CBAM & 618,013 &       & 0.626 & 0.530 & 0.919 & 0.659 & \textbf{0.960} & 0.954 \\
    ResNet-54 & 1,666,145 &       & 0.609 & 0.527 & 0.918 & 0.642 & \textbf{0.960} & 0.957 \\
    VGG16 & 2,357,329 &       & 0.617 & 0.523 & 0.920 & 0.651 & 0.948 & 0.946 \\
    ResNet-50 & 8,597,969 &       & 0.559 & 0.481 & 0.894 & 0.597 & 0.876 & 0.952 \\
    Xception & 20,843,801 &       & 0.576 & 0.503 & 0.905 & 0.612 & 0.925 & 0.946 \\
    ResNext+CBAM & 106,172,658 &       & 0.590 & 0.500 & 0.907 & 0.626 & 0.925 & 0.953 \\
    \midrule
    Sen2LCZ-Net-MF(F16-L17) + skip connection & 791,428 &       & \textbf{0.641} & \textbf{0.564} & \textbf{0.924} & \textbf{0.671} & 0.954 & \textbf{0.963} \\
    Sen2LCZ-Net-MF(F16-L17) + CBAM & 815,836 &       & \textbf{0.655} & \textbf{0.556} & \textbf{0.930} & \textbf{0.685} & \textbf{0.965} & 0.960 \\
    \bottomrule
    \end{tabular}%
  \label{tab:acc}%
\end{table*}%

\begin{figure*}
	\centering
	\includegraphics[width=0.75\textwidth]{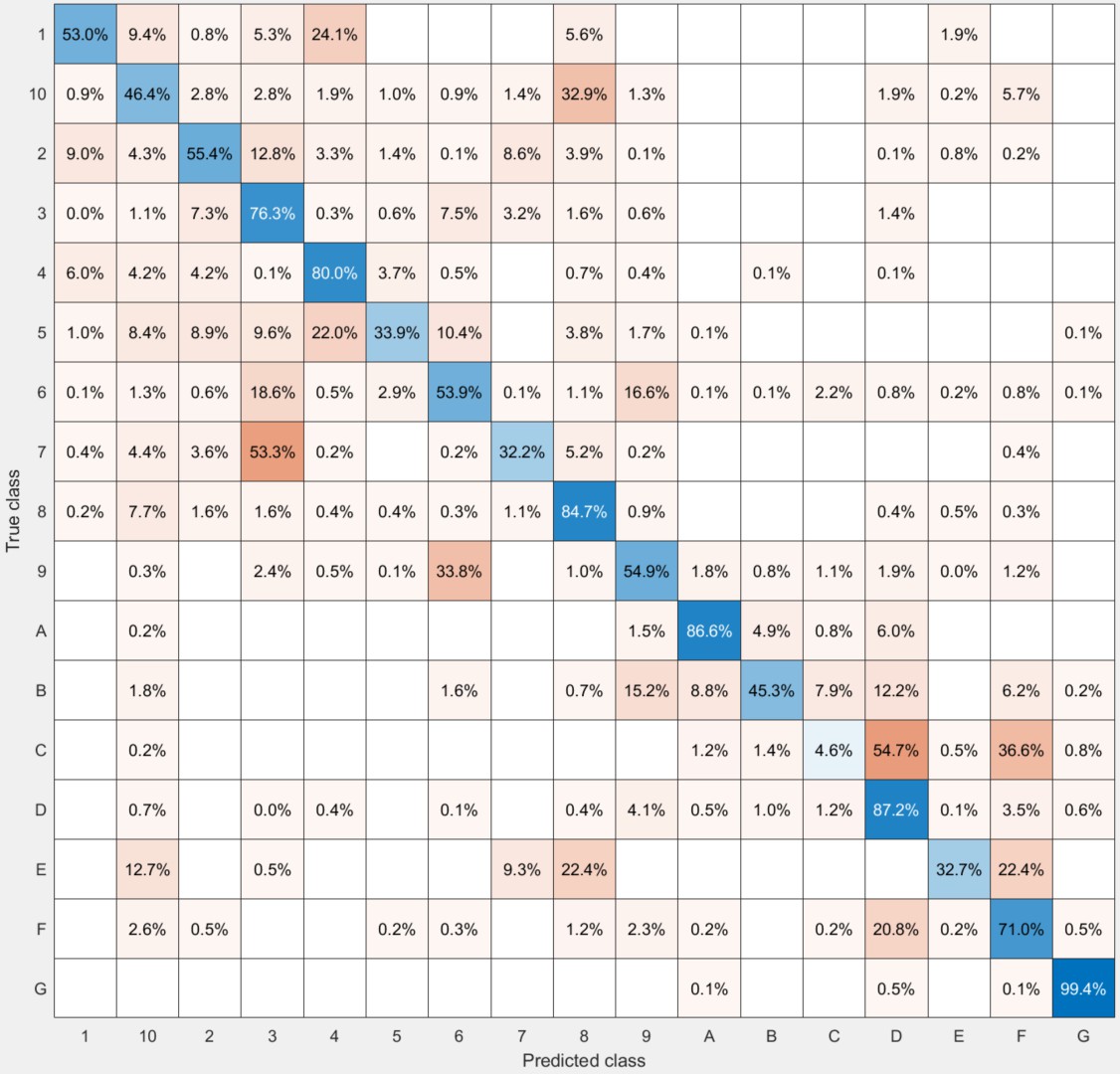}
	\caption{{Confusion matrix of classification results from Sen2LCZ-Net-MF(f16D17), normalized by the number of total samples of each LCZ.
	}}
	\label{fig:confu}
\end{figure*}

The confusion matrix resulting from Sen2LCZ-Net-MF(f16D17) is presented in Fig. \ref{fig:confu}. It can be seen that LCZs with a low producer's accuracy (lower than 50\%) include LCZ 5, 7, 10, B, C, and E. The main mis-classifications (higher than 30\%) are between LCZ 7 and 3, LCZ 9 and 6, LCZ 10 and 8, LCZ C and D, and LCZ C and F, which are all comparably similar LCZ types. The LCZs with high producer's accuracy (higher than 80\%) are LCZ 4, 8, A, D, and G.

The seemingly disappointing classification results are due to the challenging setup in our experiments\textcolor{black}{,} where the test samples are from completely unseen areas in spatially disjoint\textcolor{black}{ed} regions compared to the data in the training set. We expect that multiple approaches can be effectively employed for further improvement, which will be discussed in Section \ref{sec:dis}.

\subsection{LCZ classification examples \textcolor{black}{on a large scale}}
\begin{figure*}[htbp]
	\centering
	\includegraphics[width=0.65\textwidth]{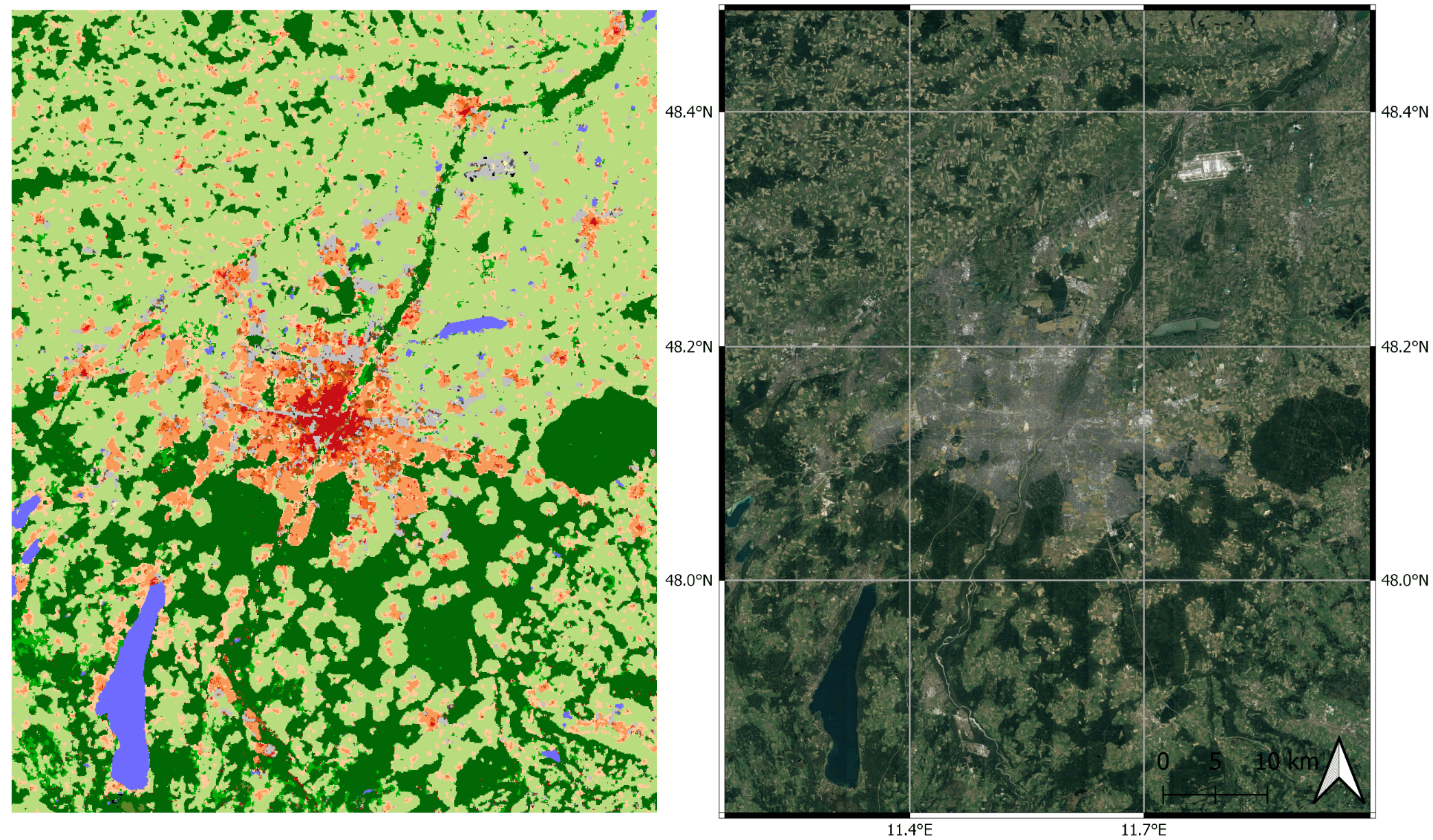}
	\caption{{LCZ classification result in Munich, Germany.
	}}
	\label{fig:munichLCZ}
\end{figure*}

\begin{figure*}[htbp]
	\centering
	\includegraphics[width=0.65\textwidth]{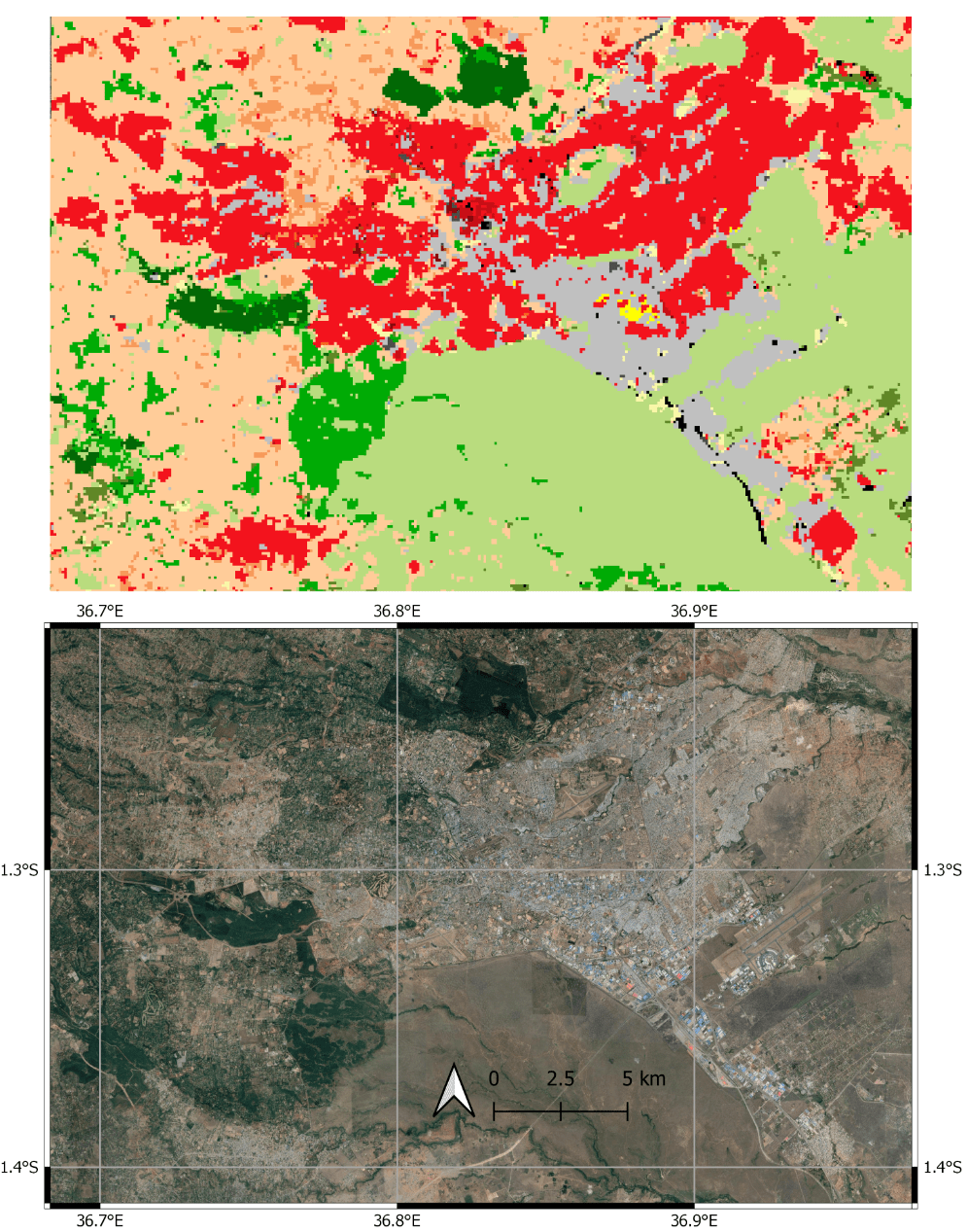}
	\caption{{LCZ classification result in Nairobi, Kenya.
	}}
	\label{fig:naiLCZ}
\end{figure*}

\begin{figure*}[htbp]
	\centering
    \includegraphics[width=0.99\textwidth]{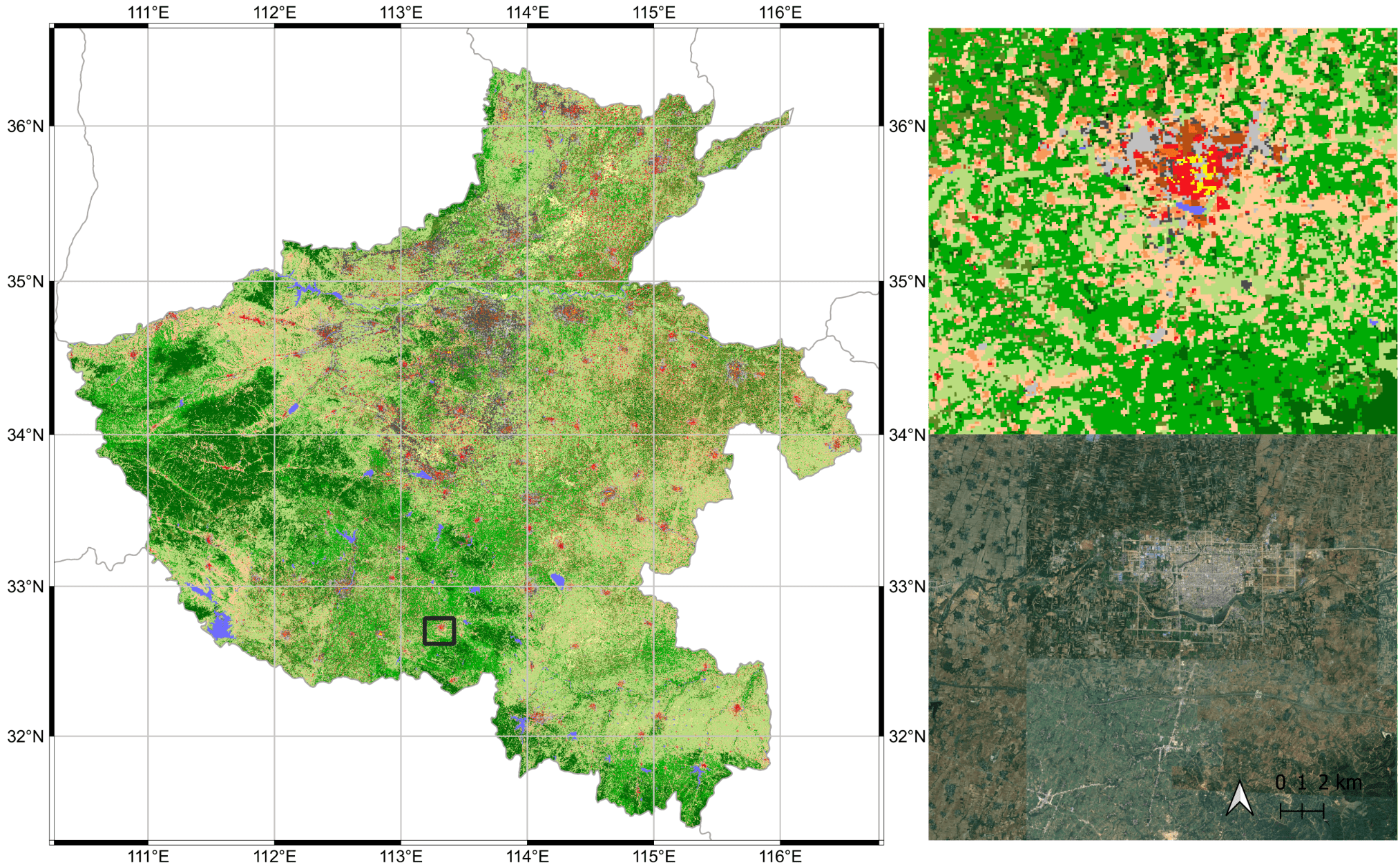}
	\caption{{An example of large-scale LCZ classification in Henan province, China, with a total area of about 167,000 $km^2$.}
}
	\label{fig:hn}
\end{figure*}

\begin{figure*}[htbp]
	\centering
    \includegraphics[width=0.99\textwidth]{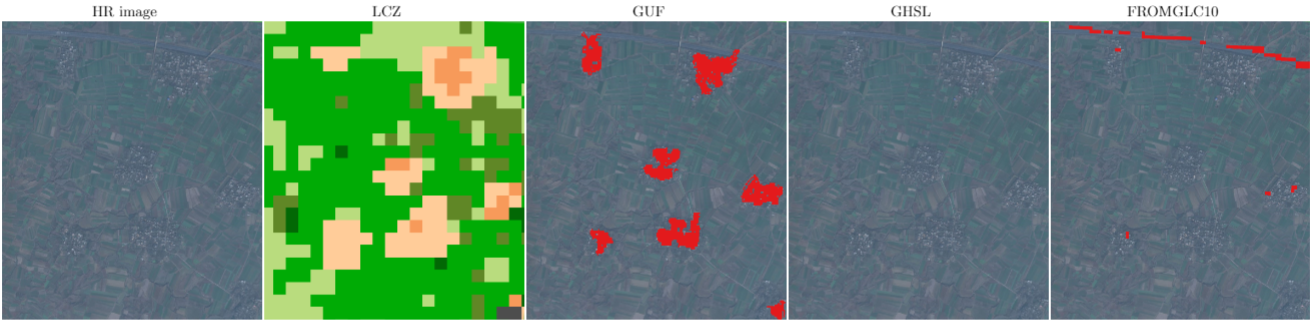}
	\caption{{Closer view of the LCZ result of large-scale classification example in Fig. \ref{fig:hn}, with an rural area around the location of longitude 113.2072\si{\degree} east and latitude 32.6849\si{\degree} north.}
}
    \begin{tikzpicture}
    \pgfplotsset{
    legend style={cells={anchor=west}, draw=none,column sep=1ex,     nodes={scale=0.7, transform shape}}
    }
        \begin{customlegend}[legend columns=8]
        \addlegendimage{water, only marks, mark=square*}
        \addlegendentry{Water}
        \addlegendimage{soil, only marks, mark=square*}
        \addlegendentry{Bare soil or sand}
        \addlegendimage{paved, only marks, mark=square*}
        \addlegendentry{Bare rock or paved}
        \addlegendimage{lowPlant, only marks, mark=square*}
        \addlegendentry{Low plants}
        \addlegendimage{bush, only marks, mark=square*}
        \addlegendentry{Bush (scrub)}
        \addlegendimage{scatTree, only marks, mark=square*}
        \addlegendentry{Scattered trees}
        \addlegendimage{denseTree, only marks, mark=square*}
        \addlegendentry{Dense trees}
        \addlegendimage{industr, only marks, mark=square*}
        \addlegendentry{Heavy industry}
        \addlegendimage{sparse, only marks, mark=square*}
        \addlegendentry{Sparsely built}

        \addlegendimage{largeLow, only marks, mark=square*}
        \addlegendentry{Large low-rise}
        \addlegendimage{light, only marks, mark=square*}
        \addlegendentry{Lightweight low-rise}  
        \addlegendimage{openLR, only marks, mark=square*}
        \addlegendentry{Open low-rise}
        \addlegendimage{openMR, only marks, mark=square*}
        \addlegendentry{Open mid-rise}
        \addlegendimage{openHR, only marks, mark=square*}
        \addlegendentry{Open high-rise}
        \addlegendimage{compLR, only marks, mark=square*}
        \addlegendentry{Compact low-rise}
        \addlegendimage{compMR, only marks, mark=square*}
        \addlegendentry{Compact mid-rise}
        \addlegendimage{compHR, only marks, mark=square*}
\addlegendentry{Compact high-rise}
\end{customlegend}
\end{tikzpicture}
	\label{fig:cmp_home}
\end{figure*}

As city-scale LCZ classification examples, we present results from Sen2LCZ-Net-MF(f16D17) for Munich (Germany, Europe) and Nairobi (Kenya, Africa) in Figs. \ref{fig:munichLCZ} and \ref{fig:naiLCZ}, respectively. As an example of province-scale LCZ classification, we present the result in Henan province in China in Fig. \ref{fig:hn}. A zoomed-in subregion of a rural area is visualized in Fig. \ref{fig:cmp_home}. \textcolor{black}{For comparative interpretation, we also present HR image data and references from existing products, i.e., Global Urban Footprint (GUF) \cite{esch2012tandem, esch2013urban}, Global Human Settlement Layer (GHSL) \cite{corbane2019automated}, and finer resolution observation and monitoring
of global land cover with 10 m GSD (FROMGLC10) \cite{gong2019stable}.}



\section{Discussion}
\label{sec:dis}

\textcolor{black}{The design choice of Sen2LCZ-Net (the suitable depth and width), the effectiveness of multi-level feature fusion and double-pooling layer, as well as the effect of class imbalance have been shown by the extensive experimental results in Section \ref{sec:res}. While having fewer parameters and not relying on sophisticated modules and training tricks, the performance of Sen2LCZ-Net-MF is beyond that of the state-of-the-art CNNs. The main reason is that most of the state-of-the-art CNNs were proposed based on the open datasets in the computer vision field, such as ImageNet and CIFAR, which are obviously different from the multi-spectral Sentinel-2 images in the So2Sat LCZ42 dataset. While some of the state-of-the-art CNNs (such as the ResNet-20+CBAM in Table \ref{tab:acc}) provide comparable results, they come with unnecessary overhead and high computation complexity. Simply relying on state-of-the-art CNNs from \textcolor{black}{the} computer vision field might be practically fast and effective for certain tasks on a small scale, but it is harmful not only for large-scale or even global mapping but also for comprehensive understanding and development of methodologies. As explained in Section \ref{sec:intro}, this is not possible without benchmark datasets, which is addressed by relying on the open So2Sat LCZ42 dataset in this study.
}


The focus of this study is to provide effective baselines on the open So2Sat LCZ42 dataset. While promising results have been achieved even in completely unseen areas, demonstrating the potential for further applications, as shown in Fig. \ref{fig:cmp_home}, our employed CNN, Sen2LCZ-Net-MF, is comparably simple and no sophisticated hyper-parameter tuning is applied on the training process, either. As a result, we expect further improvement from a range of feasible approaches based on the achieved benchmark results in this study. We categorize these possibilities in the following three directions.

\begin{itemize}
\item \textbf{Data-driven approach.} It has been shown that the performance of a CNN will increase with an increasing amount of data available \cite{simonyan2014very}. Therefore, a simple way to enhance the benchmark results is to extend the So2Sat LCZ42 dataset by including more data. A straightforward approach is to resort to data augmentation, such as the multi-scale and horizontal flip approach, and \textcolor{black}{test-time augmentation}, which has been used in state-of-the-art research \cite{simonyan2014very}. One step further, we can include more data from even more diverse areas. In addition to the amount of data, the quality of samples can also be improved to enable more accurate results. For instance, a more balanced distribution among all classes can help training a more robust model. Another example is to introduce more hard samples such as LCZ 7 (light-weight and low-rise) and LCZ 9 (sparsely-built). In this way, the learned features for hard examples can be more representative and the accuracy for difficult classes can be improved, resulting in a higher overall accuracy \cite{hughes2018mining}.

Apart from the dataset to train the network, in the prediction phase, multi-source multi-temporal data fusion is also a straightforward and effective approach to further improve the obtained benchmark LCZ classification results. The effectiveness of data fusion has been shown in \cite{qiu2019fusing}, \cite{iannelli2019urban}, and \cite{rosentreter2020towards}. Specifically, a final robust result can be achieved via a decision-level fusion of multiple predictions that are obtained from multi-source data, such as SAR and hyperspectral image, with same or different classifiers \cite{yokoya2018open, hong2019learnable, hong2019cospace, hong2020learning, hong2020invariant}. 

\item \textbf{Model-based approach.} The first kind \textcolor{black}{of model-based approaches} is ensembling of CNNs, which was observed very early and is now commonly used for obtaining top results in image classification challenges. This effectiveness is due to the presence of several local minima for the problem. Therefore, multiple trainings of the exact same neural network architecture can lead to different outputs. It should be mentioned that an ensemble does not have to \textcolor{black}{increase} the cost of computation time (training a CNN multiple times) \cite{cirecsan2012multi}. Instead, the spirit of the ensemble can be realized at different levels, for instance, by multi-column and multi-branch architectures, where the training only needs to be carried out once \cite{chollet2017xception, gastaldi2017shake, dutt2019coupled}. Even with a single-branch architecture, an ensemble can be performed with a special training strategy that passes a range of local minima during the training process \cite{huang2017snapshot}.

More interestingly, advanced algorithms in transfer learning, active learning, and meta learning can also be adapted for the LCZ classification task on the So2Sat LCZ42 dataset. For example, one challenge in large-scale LCZ classification is to achieve reasonable results in areas where no or only little reference data is available. In this case, zero-shot and few-shot learning based on meta learning principles are very promising directions to explore \cite{finn2017model, kemker2018low}.

\item \textbf{Application-oriented approach.} Depending on the specific cases of application, this study can be extended further. For instance, when an LCZ map of a certain region is required for a surface urban heat island (SUHI) study, pre-trained models can be fine-tuned on the available reference data of this area. In this way, better results can be obtained to satisfy the application\textcolor{black}{,} even though the model might overfit to this specific area. Another use case is monitoring of urbanization, which is increasingly attracting attention. In this case, the urban-related LCZs can be combined and multiple predictions from time series remote sensing images can be obtained for a post-classification change detection and analysis \cite{vandamme2019revealing}.

\end{itemize}


\section{Conclusions and Outlook}
\label{sec:con}

A range of benchmark results on the open So2Sat LCZ42 dataset were presented in this study. Because of the consistent experimental setup, this work can enable a complete understanding of the performance of CNNs on a large-scale LCZ classification task. We show that a properly designed simple CNN considering multi-level feature fusion can perform better than bigger and more complex models that \textcolor{black}{have been} proposed on non-remote sensing datasets. Since the proposed model is simple and light weight, further accuracy improvement within a certain budget on the model size or memory can be expected. As is well known, ultimately it’s a problem of well-balanced compromise between performance and imposed overhead for the specific use cases. Furthermore, our work will facilitate the development of more advanced models for the challenging task of LCZ classification \textcolor{black}{on a large scale}. Our trained models can be used as pre-trained ones, either using the fixed So2Sat LCZ42 features or fine-tuning from the So2Sat LCZ42 initialization, for related studies such as land cover land use classification from Sentinel-2 or Landsat data.


\section{ACKNOWLEDGEMENTS}

The work is jointly supported by the China Scholarship Council (CSC), the European Research Council (ERC) under the European Union's Horizon 2020 research and innovation programme (grant agreement No. [ERC-2016-StG-714087], Acronym: \textit{So2Sat}), by the Helmholtz Association
through the Framework of Helmholtz Artificial Intelligence Cooperation Unit (HAICU) - Local Unit ``Munich Unit @Aeronautics, Space and Transport (MASTr)'' and Helmholtz Excellent Professorship ``Data Science in Earth Observation - Big Data Fusion for Urban Research'' and by the German Federal Ministry of Education and Research (BMBF) in the framework of the international future AI lab "AI4EO -- Artificial Intelligence for Earth Observation: Reasoning, Uncertainties, Ethics and Beyond". The work of B. Bechtel is supported by DFG Project 437467569. 

\bibliographystyle{ieeetr}
\bibliography{ref.bib}

\end{document}